**Distribution Systems Study Group-GDI**

**Effects of Distributed Generation on the Bidirectional Operation of Cascaded Step Voltage Regulators: Case Study of a Real 34.5 kV Distribution Feeder**

HUGO RODRIGUES DE BRITO(1)*; VALÉRIA MONTEIRO DE SOUZA(1); JOÃO PAULO ABREU VIEIRA(1); MARIA EMÍLIA DE LIMA TOSTES(1); UBIRATAN HOLANDA BEZERRA(1); VANDERSON CARVALHO DE SOUZA(1); DANIEL DA CONCEIÇÃO PINHEIRO(1); HEITOR ALVES BARATA(1); HUGO NAZARENO DE SOUZA CARDOSO(2); MARCELO SOUSA COSTA(2);
FEDERAL UNIVERSITY OF PARÁ(1);EQUATORIAL ENERGIA UTILITY COMPANY(2);

ABSTRACT

This work investigates the impact of feeder bidirectional active power flow on the operation of two cascaded step voltage regulators (SVRs) located at a 34.5 kV rural distribution feeder. It shows that, when active power flow reversal is possible both by network reconfiguration and by high penetration levels of distributed generation (DG), typical SVR control mode settings are unable to prevent the occurrence of runaway condition, a phenomenon characterized by loss of SVR voltage control capabilities. Such developments are the basis for a DG pre-dispatch control strategy that aims to avoid the adverse effects of the described power flow reversal scenarios, as well as to ensure reliable operation of the utility distribution network.

1.0 - INTRODUCTION

Modern medium-voltage distribution grids are rapidly developing and adapting in response to emerging planning and operation trends, especially with the ever-increasing number of distributed generation (DG) grid interconnections requested by independent power producers (IPPs). Arising scenarios of bidirectional active power flow and short-circuit current lead to previously atypical technical issues, such as overvoltage problems along the feeders (1). Thus, the adequate operation of autonomous voltage control devices – e.g. step voltage regulators (SVRs) – is paramount for maintaining steady-state line voltages within acceptable operating ranges, as per local regulatory standards. Managing these devices is one of the biggest current challenges faced by Brazilian electric utilities.

In Brazil, cascade connections of SVRs are typically employed in medium-voltage radial distribution feeders, which cover long geographical distances in predominantly rural and sparsely populated areas (2). Such devices comprise two basic control mode settings, suitable for different operational scenarios: bidirectional mode, if active power flow reversal is possible due to a network reconfiguration scheme via switching operations with neighboring feeders; and cogeneration mode, if active power flow reversal is possible due to the presence of a high-capacity DG unit whose penetration level exceeds feeder loading downstream of the SVRs. In case these recommendations are not followed or both scenarios of active power flow reversal are simultaneously possible, the SVRs become susceptible to a circumstance of voltage control capability impairment known in literature as reverse power tap changer runaway condition, or simply SVR runaway condition (3). Consequences of this phenomenon include sustained overvoltage or undervoltage levels throughout the feeder, as well as higher maintenance costs, useful life reduction and excessive wear and tear of the affected equipment.

*Perimetral Avenue, n° 2651 – Building 01 – Center of Excellence in Energy Efficiency of the Amazon Region (CEAMAZON/UFPA) – Laboratory of Power Systems Modelling and Analysis – P.C. 66.077-830 Belém, PA, Brazil
Tel: (+55 91) 98048-0614 – E-mail: hugodbrito93@gmail.com

The runaway condition of multiple bidirectional SVRs subjected to DG-caused reverse active power flow is detailed in (2), (3). On the other hand, the authors of (4), (5) investigate active power flow reversal due to load transfer, among other network reconfiguration schemes, in order to propose centralized control strategies capable of eliminating risks related to such operations while maintaining grid reliability. In (6), a similar problem is tackled using a decentralized solution, but SVR runaway condition due to high DG penetration is not addressed. Works that simultaneously cover both active power flow reversal possibilities are noticeably scarse in literature, and those that do generally resort to control strategies based on heavy investments in communication links between devices. The authors of (7) fill this gap by presenting an alternative operational setting for the SVRs, designed with the specific goal of mitigating the aforementioned issue, but such novel functionality is not currently implemented in local SVR controllers, as can be seen in manuals for typical Brazilian models such as (8).

This work presents a solution for current local utility demands regarding a mitigation strategy to prevent the occurrence of SVR runaway condition in a 34.5 kV rural distribution feeder containing two SVRs in cascade connection. Case studies are conducted via quasi-static time series (QSTS) simulations in the Open Distribution Simulator Software (OpenDSS). Using real system parameter data, it is shown that active power flow reversal is possible not only by switching operations with a neighboring feeder, but also by a high-capacity synchronous machine-based DG unit located at the far-end of the feeder. Typical SVR control mode settings and load demand profiles are considered as a means to establish and support the proposed DG pre-dispatch control strategy, which is simple to implement and ensures reliable operation of the utility distribution network.

## 2.0 - GENERAL CHARACTERISTICS OF STEP VOLTAGE REGULATORS

The SVR device conventionally employed in Brazilian medium-voltage distribution grids is essentially an auto-transformer with a load tap changing mechanism in its series winding. Depending on feeder loading, the voltage induced at such winding can either be added to or subtracted from the primary voltage, which allows for bilateral adjustments of small deviations with respect to a user-defined voltage setpoint. Many SVR models present a total regulation range of ±10 %, equally divided into 33 discrete steps: 16 voltage raising positions, 16 voltage lowering positions, and the neutral position. Therefore, each tap operation amounts to a 0.625 %, or 0.00625 normalized p.u., change per step. Details concerning constructive aspects of the single-phase SVR used in this work's simulations are available in (8). The following subsections summarize the main parameters and technical requisites for the proper operation of SVRs installed at strategic points along a feeder.

### 2.1 - Simplified SVR control system model

Figure 1 shows the simplified control system model of an SVR. The input voltage ($V_{in}$) measured at the SVR regulation point – either its load or its source terminal, depending on operation settings – is compared to the user-defined voltage setpoint ($V_{ref}$), thus resulting in a voltage error signal ($V_{error}$) sent to the "Measuring Element" block. The line drop compensation (LDC) feature, which allows for $V_{in}$ to be remotely estimated at a point farther from SVR terminals, is not considered in this work because it is not within the local utility standard operation procedures.

The voltage error signal is in turn compared to the deadband (D), an adjustable range of allowed variance around $V_{ref}$, and at times to the hysteresis band ($\varepsilon$), a parameter that mitigates frequent tap operations during temporary oscillations around D. If $V_{error}$ exceeds such limits, an activation signal ($V_{act}$) is sent to the "Tap Changer" block, triggering its timer relay. Although time delay schemes vary widely depending on Brazilian utility practices, several of them employ a double-time delay scheme, where the first tap operation trigger ($T_1$) is slower than the subsequent ones ($T_2 = T_3 = \ldots$). Once the relay times out, the tap changing command is sent to a motor drive unit, which mechanically carries out the operation. This procedure is repeated as many times as necessary until $V_{error}$ is back within the predetermined deadband limits.

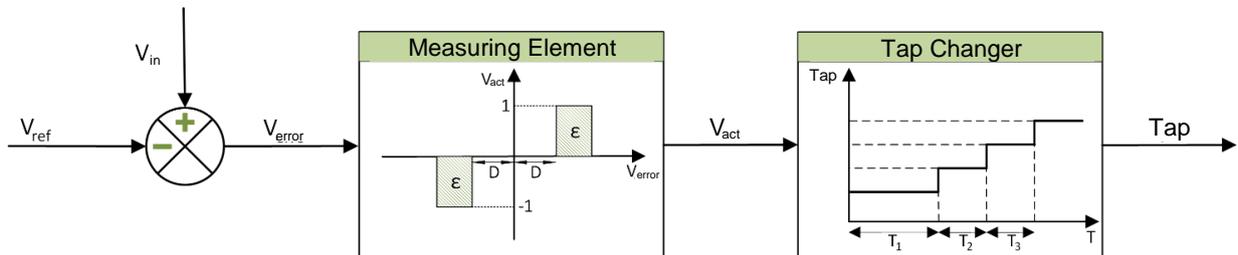

FIGURE 1 - Simplified control system model of a step voltage regulator.

### 2.2 - Operation in cascade connection

The cascade or series connection of SVRs is defined as the installation of two or more of these devices in the main branch of the same feeder, requiring special time delay adjustments so as to avoid unnecessary tap operations. In (9), a limit of four SVRs installed in cascade connection is established for Brazilian distribution feeders.

With respect to the double-time delay scheme of Figure 1, the $T_1$ parameter of the SVR closest to the primary substation (PS) must be the smallest of the cascade connection, and increase progressively as subsequent SVRs are located farther and farther from the PS. This configuration allows for the most influential SVR throughout the feeder to have a faster response to voltage variations, effectively reducing the number of tap operations of downstream SVRs. Such adjustment is only valid for $T_1$, since subsequent tap operations are usually set to a constant time delay of 5 s for all SVRs. Figure 2 shows a properly adjusted cascade connection of four SVRs, with typical first tap operation delays. It is worth noting that SVRs located at side branches or at neighboring feeders connected via switching operations are not to be considered for the delay adjustment scheme of the main branch, which in this case is characterized by the gradual $T_1$ increase of 15 s per SVR.

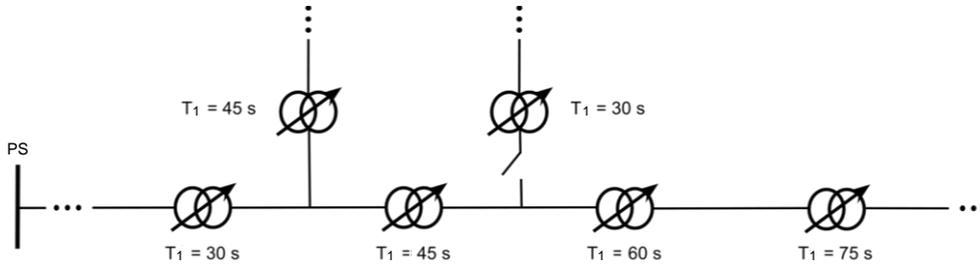

FIGURE 2 - Feeder containing four cascaded SVRs along its main branch.

## 3.0 - CONTROL MODE SETTINGS OF STEP VOLTAGE REGULATORS

### 3.1 - Bidirectional mode

In bidirectional mode, the SVR determines its regulation point based on the direction of the active power flow. Figure 3 depicts the situation of direct active power flow through the feeder, when the DG supplies less real power than the load center downstream of the SVR demands. In this case, the resulting active power flows through the SVR from PS to DG, and the device regulates its load terminal (point 2), located on the lower short-circuit capacity side of the feeder. This operational scenario is considered acceptable, since voltage control through tap changing is effective.

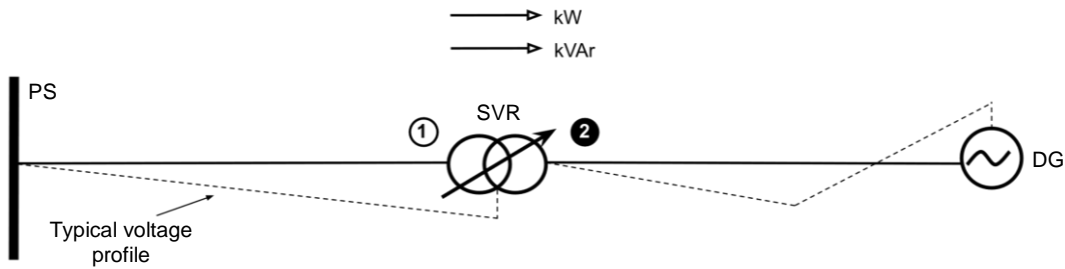

FIGURE 3 - Bidirectional SVR during direct active power flow through the feeder.

Figure 4 depicts the situation of reverse active power flow through the feeder, when the real power supplied by the DG exceeds the load center demand. In this case, the resulting active power through the SVR flows from DG to PS, and the device regulates its source terminal (point 1), located on the higher short-circuit capacity side of the feeder. Therefore, the tap changer operates in an effort to reduce the voltage at this point with negligible results, given the electrically strong nature of the upstream portion of the feeder. The outcome is a sequence of failed regulation attempts and, due to the ensuing reactive power flow, a net effect of significant voltage rise at point 2. Successive operations continue until the tap limit is reached, leading to a 10 % overvoltage downstream of the SVR.

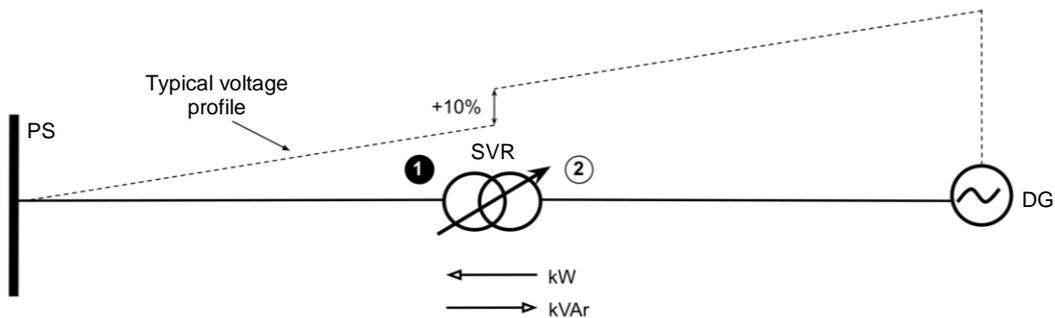

FIGURE 4 - Bidirectional SVR during reverse active power flow through the feeder.

The loss of SVR voltage control capability by attempting to regulate a point on the higher short-circuit capacity side of the feeder characterizes SVR runaway condition. The possibility of reverse active power flow through the SVR due to high DG penetration levels, even if temporarily, makes SVR bidirectional mode unacceptable in a real-world scenario. Moreover, the authors of (2) show that cascade connections of bidirectional SVRs might lead to a cumulative effect of voltage violation (usually overvoltage) when the runaway phenomenon occurs, further aggravating the issue throughout the feeder, and especially at its far-end.

## 3.2 - Cogeneration mode

In cogeneration mode, the SVR does not take the active power flow direction into consideration when determining the regulation point, which is maintained invariably at its load terminal (point 2). Figure 5(a) depicts both situations of direct and reverse active power flow presented at Figure 3 and Figure 4, respectively, now with the SVR set to cogeneration mode. In these circumstances, this control mode setting allows for effective SVR operation always on the lower short-circuit capacity side of the feeder, preventing the occurrence of runaway condition and ensuring an acceptable operational scenario.

However, in the event of network reconfiguration via switching operations with a neighboring feeder, the regulation point of the SVR remains the same, but now becomes part of the higher short-circuit capacity side of the reconfigured feeder, as shown in Figure 5(b). Thus, the device attempts to regulate the electrically strong upstream portion of the reconfigured feeder, to little avail. The resulting loss of voltage control capability indicates SVR runaway condition, as well as other control issues if the neighboring feeder contains an SVR set to regulate the same area as the original SVR, as is the case with SVR1 and SVR2 in Figure 5(b). Hence, the possibility of reverse active power flow through the SVR due to network reconfiguration makes cogeneration mode unacceptable in a real-world scenario.

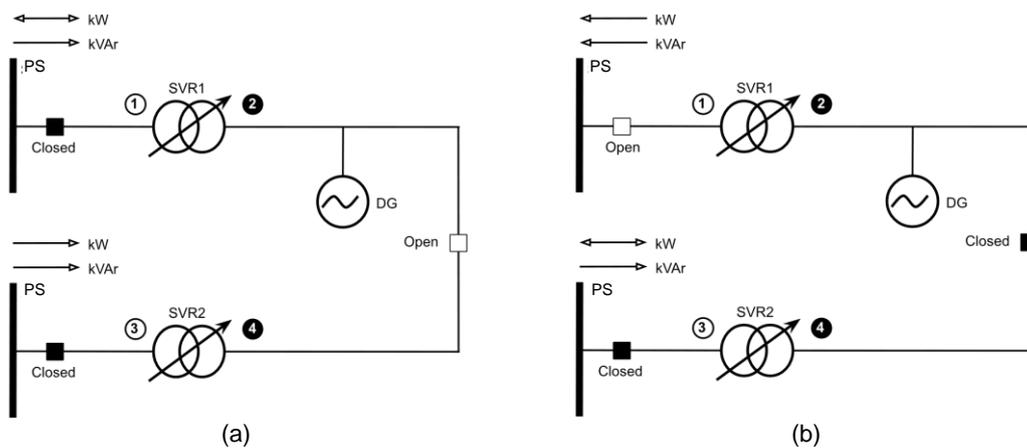

(a) (b)

FIGURE 5 - SVR cogeneration mode (a) before network reconfiguration; (b) after network reconfiguration.

## 4.0 - RESULTS

At the Brazilian substation of Paragominas, there are two interconnected rural distribution feeders, named PR-09 and PR-11, whose operations are a responsibility of the local Pará state utility. Both have a base voltage of 34.5 kV and main branches of roughly 100 km of length. The feeders present radial topology and their lines are mainly comprised of the following power cables: 2 AWG ACSR, 1/0 AWG ACSR, 336 MCM and 4/0 AWG ACSR. The PS has a Thévenin impedance of $Z_0 = (0.002 + j12.13)\ \Omega$ and $Z_1 = (1.369 + j17.633)\ \Omega$. To ensure adequate voltage levels at the far-end of both feeders, each of them contains two SVRs in cascade connection with a double-time delay scheme. First tap operation delays assigned to the SVRs closest to the PS and to the ones farther downstream in each feeder are of, respectively, 30 s and 45 s. Subsequent tap operation delays are in all devices equal to 5 s. Figure 6 shows the single-line diagram of the PR-09 and PR-11 feeders, along with the location of certain points of interest for this work.

The Open/Closed (Open = 0, Closed = 1) settings of the circuit breakers of PR-09 (CB1) and PR-11 (CB2), the switch located at PR-09 (SW1) and the switch that connects both feeders (SW2) enable four different operational scenarios. If CB1 = CB2 = SW1 = 1 and SW2 = 0, both feeders operate independently of each other. If CB1 = SW1 = SW2 = 1 and CB2 = 0, PR-09 becomes responsible for both feeders' full load demand. Conversely, if CB2 = SW1 = SW2 = 1 and CB1 = 0, PR-11 becomes responsible for both feeders' full load demand instead. The final scenario corresponds to the partial transfer of a fraction of PR-09 load to PR-11, thereby requiring CB1 = CB2 = SW2 = 1 and SW1 = 0.

A large industrial plant that produces ethanol and sugar, currently off-grid, is expected to be connected to bus 1050 of PR-09. After such interconnection, the industry will operate as an IPP and its 12.5 MVA synchronous machine will work as the IPP's DG, contributing to the network's total three-phase short-circuit current with 632.05 A. According to the contract signed with the utility, the DG, set to constant unity power factor mode, will supply the total industry load (4.3 MW + j1.83 MVAr) and inject uninterrupted 3 MW into the grid for 6 months in a row, during sugar cane harvest.

To investigate possible issues related to the IPP integration to PR-09, several simulations were conducted considering the aforementioned four operational scenarios. QSTS simulations performed in OpenDSS have a time step of 1 s and total simulation time of 24 h, with the exception of shorter simulations aimed at detailing SVR runaway progression over time, which only last 250 s. Load demand profiles and SVR settings are in accordance with typical utility practices. Total feeder real power demand varies within the range of 2.13 MW to 2.55 MW for PR-09, excluding the industry load, and of 1.22 MW to 4.03 MW for PR-11. Thus, DG penetration level might reach 172.89 % if connected to PR-09 and 118.03 % if connected to PR-11 via network reconfiguration.

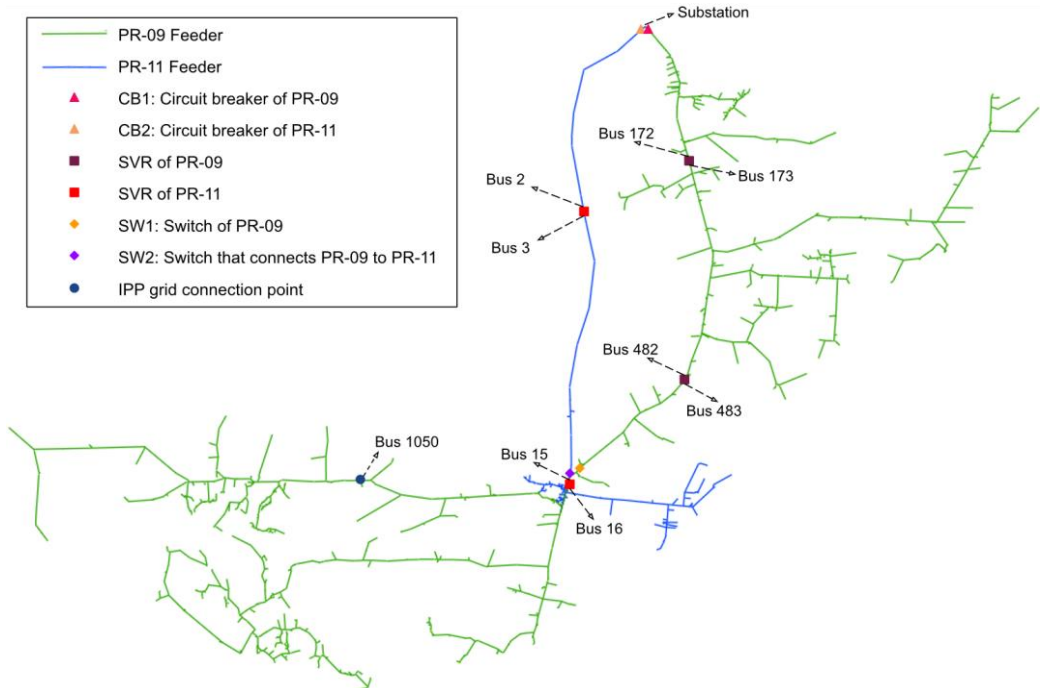

FIGURE 6 - Single-line diagram of the PR-09 and PR-11 feeders.

## 4.1 - PR-09 operation with disconnected SVRs

Based on contractual requirements and feeder demand profiles, it is evident that the DG injection exceeds PR-09 full loading even during its heaviest demand hours. It follows that the SVRs might no longer be necessary to regulate voltage drop along the feeder. By disconnecting them, runaway events could be avoided and all four operational scenarios could be reliably employed to improve system operation. A simulation of selected PR-09 bus voltages following this logic is shown in Figure 7, for the SW2 = 0 scenario. As it can be observed, interconnecting the IPP's generator to PR-09 with disconnected SVRs leads to severe overvoltage problems, especially at bus 1050 and at the IPP bus, sustained throughout the day. Thus, the SVRs are necessary to mitigate voltage rise after DG integration.

Alternatively, the DG could be set to voltage control mode, which is already considered by (10) as an admissible DG configuration upon agreement with the utility. However, since significant reactive power exchanges would be required to mitigate the feeder overvoltage, the DG power factor would drop below the 0.92 lower limit and total system losses would increase substantially as shown in (2), making this solution impractical from a techno-economic standpoint.

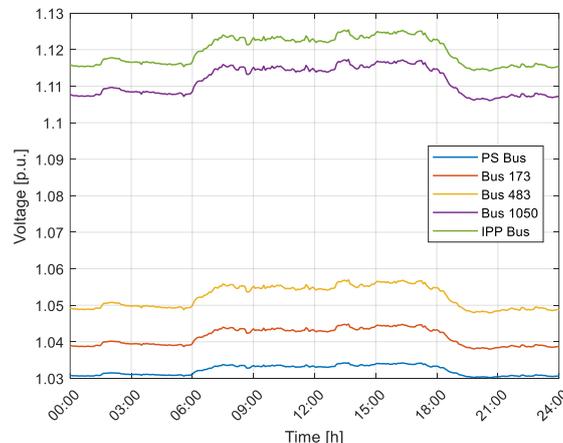

FIGURE 7 - Selected PR-09 bus voltages for SW2 = 0.

4.2 - <u>PR-09 and PR-11 operation with SVRs in cogeneration mode</u>

Simulations performed in this subsection tackle SVR behavior after DG interconnection. This investigation aims to assess the devices' performance in mitigating the previously observed feeder overvoltage. As such, PR-09 is set to operate independently (SW2 = 0) and both SVRs are adjusted to 0.97 p.u. voltage setpoint, 1 % deadband and no hysteresis band. Additionally, since DG active power injection always exceeds PR-09 full loading, both SVRs are set to cogeneration mode, which prevents DG-caused runaway condition. Figure 8 depicts the results obtained.

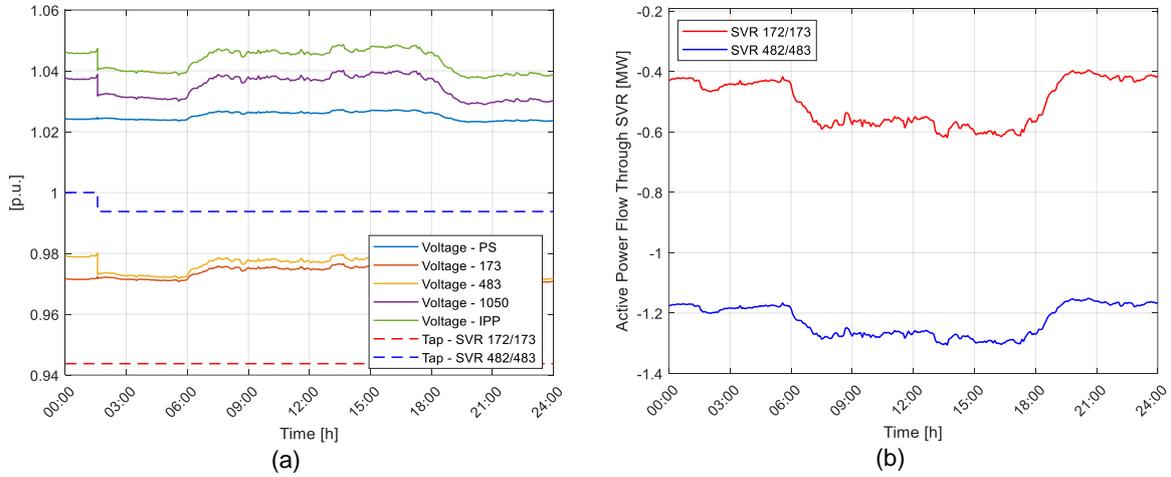

(a)                                    (b)

FIGURE 8 - (a) PR-09 bus voltages and SVR tap operations for SW2 = 0; (b) Active power flow through the SVRs.

Figure 8(a) shows the same selection of bus voltages as Figure 7, now within adequate operating ranges, along with the few SVR tap operations required to correct deadband violations at their respective regulation points. Thanks to such corrections, this operational scenario becomes acceptable. Figure 8(b) shows that reverse active power flow through both SVRs is maintained throughout the day. Thus, the bidirectional mode setting, necessary for CB1 = 0 and CB2 = 0 network reconfiguration schemes, is not applicable in this case owing to the subsequent SVR runaway phenomenon occurrence, as explained in Subsection 3.1.

An alternative solution requires grid topology modification via partial PR-09 load transfer to PR-11, i.e., the SW1 = 0 scenario. Keeping the remaining switch and circuit breakers closed, the far-end portion of PR-09, IPP included, is easily transferred over to PR-11. This operational scenario only affects the behavior of the PR-11 SVR closest to the PS, located between bus 2 and bus 3 (SVR 2/3), since its other SVR (SVR 15/16) is located downstream of SW1. An advantage of this setting is that, the PR-11 demand being heavier than the PR-09 demand on average, the magnitude of overvoltage in the reconfigured feeder is naturally reduced. To this end, both SVRs are adjusted to 1 % deadband, no hysteresis band and different voltage setpoints: 0.975 p.u., for SVR 2/3, and 1 p.u., for SVR 15/16. Similar to the previous SW2 = 0 case, both SVRs are set to cogeneration mode because, even with heavier feeder loading, the DG might still cause active power flow reversal through SVR 2/3 (but not through SVR 15/16, again due to its location) during medium and light demand hours. Figure 9 depicts the simulation results obtained.

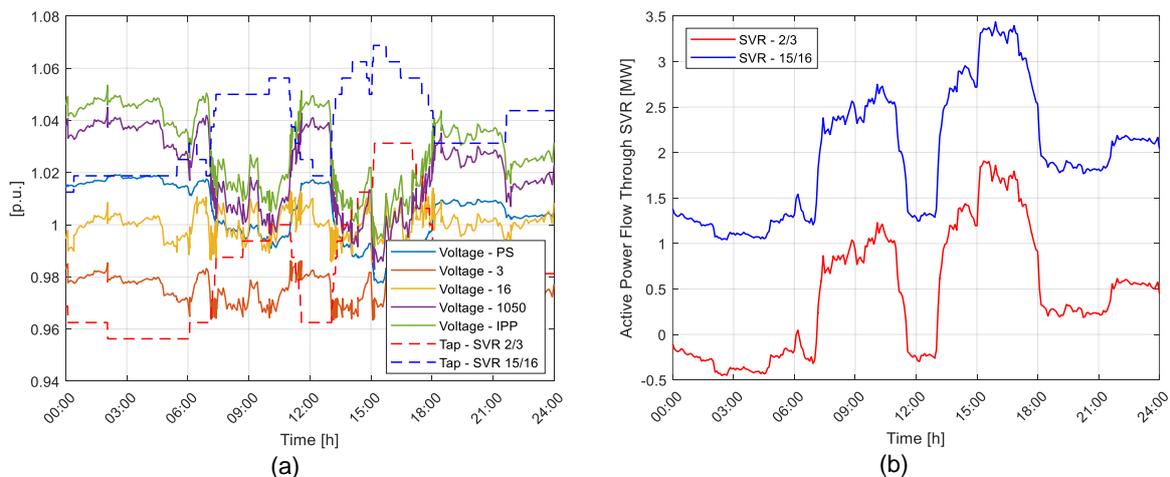

(a)                                    (b)

FIGURE 9 - (a) PR-11 bus voltages and SVR tap operations for SW1 = 0; (b) Active power flow through the SVRs.

Figure 9(a) shows that bus voltages are within normal operating ranges, thanks to effective SVR tap operations. Once again, this scenario is considered acceptable. Figure 9(b) shows that active power flow through SVR 15/16 is always direct as expected, whereas reverse active power flow through SVR 2/3 occurs during light and medium demand hours. Thus, setting SVR 2/3 to bidirectional mode is not possible for the same reasons as in the SW2 = 0 case.

Although both solutions provide results in which the feeder operates within adequate voltage deadband limits, they share the disadvantage mentioned in Subsection 3.2 with respect to network reliability reduction. If PR-09 operates independently (SW2 = 0), PR-11 will not be able to temporarily assume both feeders' full load demand (CB1 = 0) in case the need arises. Similarly, if a fraction of PR-09 load is transferred over to PR-11 (SW1 = 0), PR-09 will not be able to temporarily assume both feeders' full load demand (CB2 = 0) in case the need arises. This is due to the fact that the affected SVRs must always be set to cogeneration mode so as to avoid the possibility of DG-caused runaway condition, being unable to change to bidirectional mode during the mentioned network reconfiguration actions.

Figure 10 depicts the circuit breaker-related scenarios to prove that SVR cogeneration mode quickly leads to runaway condition after network reconfiguration. Figure 10(a) starts with SW2 = 0 until the 20 s mark, when CB1 = 0 is triggered and PR-09 SVRs runaway soon follows, as evidenced by the step-by-step voltage rise progression at downstream buses. A cumulative overvoltage effect arising from both SVRs' runaway can also be observed specifically at bus 172, due to its location in the reconfigured feeder. Figure 10(b) starts with SW1 = 0 until the 20 s mark, when CB2 = 0 is triggered and PR-11 SVR 2/3 runaway soon follows, this time showing that undervoltage is also a possible outcome.

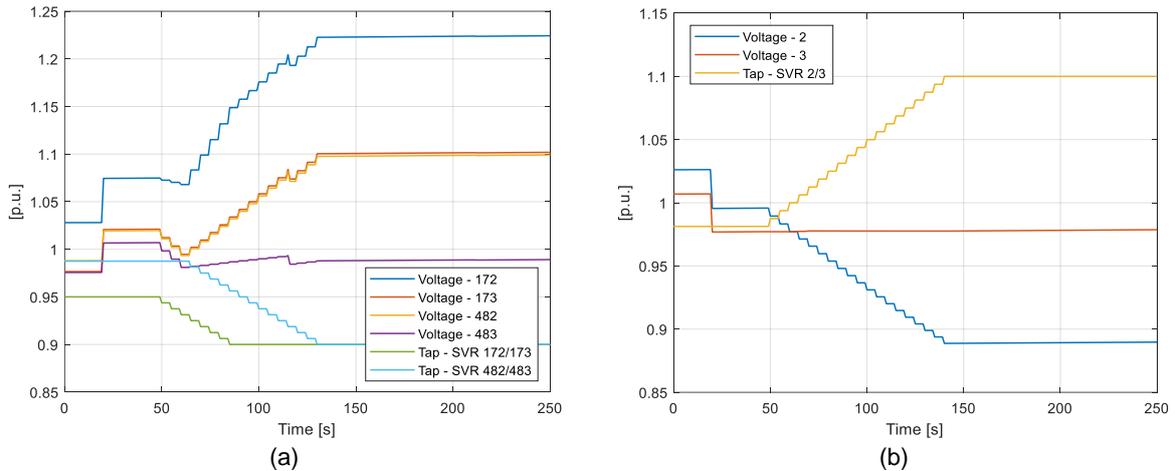

(a)                                          (b)

FIGURE 10 - Runaway condition in (a) PR-09 SVRs for CB1 = 0; (b) PR-11 SVR 2/3 for CB2 = 0.

### 4.3 - PR-11 operation with SVRs in bidirectional mode and the proposed DG pre-dispatch control strategy

To circumvent the aforementioned issues, a DG pre-dispatch control is proposed to improve the SW1 = 0 scenario reliability, with the IPP transferred over to PR-11. The solution consists in adjusting, via ramp variations, the daily DG active power injection according to feeder load forecast data provided by the utility. That way, the DG is scheduled to reduce its real power dispatch during light demand hours and increase it during heavy demand hours. Adjustments are made in such a way that the daily average DG active power injection is equal to 3 MW, which is an acceptable equivalent to the uninterrupted 3 MW contractual requirement. The pre-dispatch rules out the possibility of DG-caused runaway condition in SVR 2/3, meaning that the device can be safely set to bidirectional mode. Thus, network reliability is improved the CB2 = 0 scenario is now applicable if needed. Figure 11 depicts results obtained from the strategy.

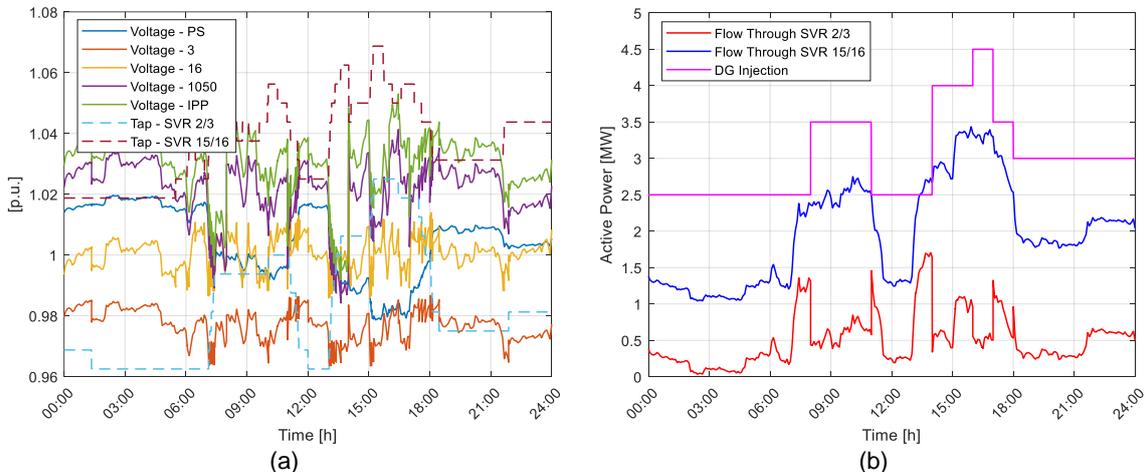

(a)                                          (b)

FIGURE 11 - (a) PR-11 bus voltages and SVR tap operations for SW1 = 0; (b) Active power flow through the SVRs and daily DG active power injection control strategy (3 MW on average).

Figure 11(a) shows that the solution ensures an acceptable scenario in terms of bus voltage profiles and SVR tap operations. Figure 11(b) shows that active power flow through the SVRs is always direct throughout the day and that the DG real power dispatch responds to feeder demand variations as proposed. The simple implementation of this strategy is made possible by the dispatchable nature of the IPP's high-capacity, synchronous machine-based DG. Under these conditions, the industrial plant can be safely interconnected into the grid as current utility challenges concerning SVR runaway condition are thereby avoided, with no reduction in existing network reliability options.

## 5.0 - CONCLUSIONS

This work investigated the impact of a high-capacity dispatchable DG on the voltage control of a 34.5 kV distribution feeder which contains cascaded SVRs and presents the possibility of network reconfiguration via switching operations with a neighboring feeder. It showed that reverse active power flow scenarios introduce new technical challenges to the conventional distribution grid, such as SVR runaway condition events, thereby increasing operational complexity. Case studies conducted via QSTS simulations in OpenDSS, using real feeder data made available by the local utility, provided the basis for the proposed DG pre-dispatch control strategy, which is tailored to current utility demands concerning the future interconnection of a large industrial plant at the far-end portion of the feeder.

The developed solution takes advantage of the IPP's high-capacity dispatchable DG to adjust its own real power injection according to feeder load forecast, via ramp-up or ramp-down variations, in such a way that the daily average active power generation is equivalent to predetermined contractual requirements. This procedure prevents the DG from reversing active power flow through the cascaded SVRs, allowing these devices to be safely set to bidirectional mode regardless of the operational scenario, thereby eliminating risks of runaway condition-related adverse effects. It is a decentralized, cost-effective and simple-to-implement strategy adequate to the Brazilian context of medium-voltage long and radial distribution feeders covering large geographical extensions, where centralized or coordinated control solutions are impractical due to the heavy investments required in communication infrastructure.

Future developments related to this work include an optimal power flow-based DG pre-dispatch method to improve control over active power ramp adjustments using different feeder demand profiles for specific days of the week and weekends. Technical constraints regarding generator limits and ramp variation rate must also factor in the formulation.

## 6.0 - REFERENCES

(1) WALLING, R.A., SAINT, R., DUGAN, R.C., BURKE, J., KOJOVIC, L. A. Summary of distributed resources impact on power delivery systems; IEEE Transactions on Power Delivery, vol. 23, no. 3, pp. 1636-1644, Jul. 2008.

(2) BRITO, H.R., SOUZA, V.M., SOUZA, V.C., VIEIRA, J.P.A., BEZERRA, U.H., TOSTES, M.E.L., GARCIA, A.O.R., COSTA, M.S., CARRERA, G.T., CARDOSO, H.N.S. Impact of distributed generation on distribution systems with cascaded bidirectional step voltage regulators; International Conference on Industry Applications, São Paulo, 2018.

(3) AGALGAONKAR, A.P., PAL, B.C., JABR, R.A. Distribution voltage control considering the impact of PV generation on tap changers and autonomous regulators; IEEE Transactions on Power Systems, vol. 29, no. 1, pp. 182-192, Jan. 2014.

(4) RANAMUKA, D., AGALGAONKAR, A.P., MUTTAQI, K.M. Online voltage control in distribution systems with multiple voltage regulating devices; IEEE Transactions on Sustainable Energy, vol. 5, no. 2, pp. 617-628, Apr. 2014.

(5) RANAMUKA, D., AGALGAONKAR, A.P., MUTTAQI, K.M. Innovative Volt/VAr control philosophy for future distribution systems embedded with voltage-regulating devices and distributed renewable energy resources; IEEE Systems Journal, in press, 2019.

(6) CHAMANA, M., CHOWDHURY, B.H., JAHANBAKHSH, F. Distributed control of voltage regulating devices in the presence of high PV penetration to mitigate ramp-rate issues; IEEE Transactions on Smart Grids, vol. 9, no. 2, pp. 1086-1095, Mar. 2018.

(7) BAGHERI, P., LIU, Y., XU, W., FEKADU, D. Mitigation of DER-caused over-voltage in MV distribution systems using voltage regulators; IEEE Power and Energy Technology Systems Journal, vol. 6, no. 1, pp. 1-10, Mar. 2019.

(8) ITB ELECTRICAL EQUIPMENT. Single-phase step-voltage regulator type transformer model RAV-2 with CTR-2 control; pp. 21-23, Jun. 2017, [Online]. Available: https://www.itb.ind.br

(9) BRAZILIAN ASSOCIATION OF TECHNICAL STANDARDS. Step voltage regulators: Specifications - NBR 11809; pp. 32-46, Feb. 1991.

(10) IEEE. Standard 1547 for Interconnection and Interoperability of Distributed Energy Resources with Associated Electric Power Systems Interface; pp. 36-41, Apr. 2018.